# Exploratory analysis of injury severity under different levels of driving automation (SAE Levels 2 and 4) using multi-source data

Shengxuan Ding[1]
Graduate Research Assistant
shengxuan.ding@ucf.edu

Mohamed Abdel-Aty[1]
Professor
m.aty@ucf.edu

Natalia Barbour[1]*
Assistant Professor
natalia.barbour@ucf.edu

Dongdong Wang[1]
Postdoctoral Scholar
dongdong.wang@ucf.edu

Zijin Wang[1]
Graduate Research Assistant
Zijin.wang@ucf.edu

Ou Zheng[1]
Research Scientist
Ouzheng1993@gmail.com

1. Smart and Safe Transportation Lab (SST), Department of Civil, Environmental and Construction Engineering, University of Central Florida,
12800 Pegasus Dr #211, Orlando, FL 32816, USA

*Corresponding author

July 2024

**HIGHLIGHTS**

- Sizable amount of data are collected by analyzing crash reports.
- The estimated models offer insights into various variables playing a role in the injury severity from crashes that involve vehicles equipped with automation.
- A comparison is made between the factors that impact the injury severity outcomes from crashes involving the ADAS and ADS equipped vehicles.

**ABSTRACT**

Vehicles equipped with automated driving capabilities have shown potential to improve safety and operations. Advanced driver assistance systems (ADAS) and automated driving systems (ADS) have been widely developed to support vehicular automation. Although the studies on the injury severity outcomes that involve automated vehicles are ongoing, there is limited research investigating the difference between injury severity outcomes for the ADAS and ADS equipped vehicles. To ensure a comprehensive analysis, a multi-source dataset that includes 1,001 ADAS crashes (SAE Level 2 vehicles) and 548 ADS crashes (SAE Level 4 vehicles) is used. Two random parameters multinomial logit models with heterogeneity in the means of random parameters are considered to gain a better understanding of the variables impacting the crash injury severity outcomes for the ADAS (SAE Level 2) and ADS (SAE Level 4) vehicles. It was found that while 67 percent of crashes involving the ADAS equipped vehicles in the dataset took place on a highway, 94 percent of crashes involving ADS took place in more urban settings. The model estimation results also reveal that the weather indicator, driver type indicator, differences in the system sophistication that are captured by both manufacture year and high/low mileage as well as rear and front contact indicators all play a role in the crash injury severity outcomes. The results offer an exploratory assessment of safety performance of the ADAS and ADS equipped vehicles using the real-world data and can be used by the manufacturers and other stakeholders to dictate the direction of their deployment and usage.

## 1. INTRODUCTION

Automation of systems has been undergoing rapid development and has revolutionized the transportation industry [1,2]. SAE International [1] categorized ADS into six levels. Level 0 and Level 1 features provide limited support in driving and include technologies such as steering or brake support. Level 2 indicates Advanced Driver Assistance Systems (ADAS), which are electronic systems that assist the drivers in driving and parking functions, utilizing features such as adaptive cruise control, lane departure warnings, and automatic emergency braking to enhance vehicle safety and improve driving experience [1]. Vehicles with automation Levels 3-5 do not need a driver (although for Level 3, a driver must take over if the vehicle's software requests it). Level 4 is also referred to as Automated Driving Systems (ADS), which use more advanced sensors and algorithms to control and navigate vehicles and while in some cases their operations will depend on the environment, they offer a high level of automation [1]. Though the advantages of ADS have been to some degree already demonstrated in smart transportation, many of these emerging technologies have also brought concerns particularly in the context of safety risks.

Although the testing of vehicles with driving automation is increasing, thus far they have been tested in limited environments. In California, for example, the manufacturers of vehicles equipped with ADS, are permitted to test their vehicles on public roads by obtaining permits from the CA DMV [2]. While only the approved vehicles and operators are permitted to conduct testing, the manufacturer must identify specific operational design domain of the automated vehicle in the permit application. During testing, the operators must follow the California traffic regulations, and the manufacturer is required to submit reports to the DMV that include the details of any incident, such as crash as well as disengagement reports [2]. The disengagement reports are needed when a vehicle equipped with ADS disengages from the automated mode either due to a technology failure, or in cases when the operator takes manual control of the vehicle. Disengagement reports also include the total Autonomous Vehicles Miles Traveled per year (AVMT). On March 17th, 2023, California DMV authorized the public release of 564 reports of crashes involving vehicles equipped with ADS that have occurred since 2014 [2]. While the goal of vehicles with varied levels of automation is to create a future with zero road fatalities, the limited testing conducted thus far has resulted in hundreds of crashes in only one state, raising concerns about their safe operations. Despite several frameworks, such as scenario-/functionality-based methods, the Long-tail Regularization (LoTR) framework [3,4], and the Parallel Vision Actualization System (PVAS) aiming to test the intelligence of autonomous vehicles, the existing legislation does not adequately address liability and safety issues.

To support the timely and transparent release of the records reflecting real-world crashes that involve vehicles equipped with ADS or ADAS from manufacturers and operators, the NHTSA issued a General Order in June 2021 requiring manufacturers and operators to report all crashes involving vehicles equipped with different levels of driving automation namely Level 2 ADAS as well as Level 4 ADS [5]. The General Order requires incident reports of crashes that occur on public roads in the United States. These reports contribute to gaining insights into the safe operations of vehicles equipped with ADS as well as provide crucial data necessary for research and understanding their driving capabilities in diverse environments.

In the current paper, we use the aforementioned data sources to conduct exploratory analysis of crash injury severity involving vehicles equipped with ADAS and ADS (Level 2 and Level 4). The dataset was crafted by processing and merging data sourced from several entities including the CA DMV [2], the NHTSA [5], OpenStreetMap (OSM), and the OpenWeather API. Notably, refined geospatial data from the official database based on specific queries were also extracted, which were subsequently cross-referenced with crash details on GoogleMap and OSM. A random parameters multinomial logit models and another random parameters multinomial logit models with heterogeneity in the mean of random parameters are estimated to identify factors influencing injury severity for crashes involving vehicles equipped with ADS and ADAS. The model estimation results provide insights into the risks associated with the evolving technology and can be used to improve their safety and further stimulate innovation. The results of this study aim to inform policymakers, engineers, and other stakeholders on the factors that contribute to autonomous vehicle crashes and guide the development of safer and more efficient autonomous vehicle systems.

The paper begins with a brief but comprehensive literature review that is followed by the description of data preparation. Next, the methodology of the models' estimation is explained, and finally, the results and conclusions are presented.

**2. LITERATURE REVIEW**

The vehicles with self-driving capabilities have attracted a lot of attention among the researchers, engineers, and public. Although the vehicles with automated driving have been marketed for their potential to reduce human driver costs, their effects on transportation planning are still unclear as the public has been expressing concerns with their adoption and usage [6-8]. Existing literature recognized safety as one of the most frequently identified concerns related to their acceptance [9-11]. Furthermore, a substantial part of past research focused on exploring the intended adoption of automated vehicles as well as identifying the essential barriers regarding their use. One Arizona based study found that familiarity with automated vehicles (AVs) does not necessarily improve safety perceptions and cycling near them was considered to be the most hazardous [12]. Stoma et al. developed an online survey to highlight the influence of socio-demographic characteristics and driving attitudes on safety perceptions to best advocate for the ongoing assessment of AVs as well as formulate adaptive strategies by the stakeholders [13]. Moody et al., on the other hand, explored global views on AV safety and found that although young, high-income males from urban areas are generally more optimistic, disparities in safety concerns may also depend on the context of developed or developing country [14].

To begin understanding automated vehicle crashes and their related injury severity, several steps must be taken. Data availability and management are essential in attempting to gain relevant and applicable insights. Because of the limited data availability, the studies addressing their actual real-life usage, related crashes, and users' experience have been scarce. Reliable data sources warrant accurate analysis and allow to plan for their effective and safe operation. A sizable amount of data are required to understand safety performance of vehicles equipped with automation. The California Department of Motor Vehicles is one of the most critical data sources that has been used thus far to investigate their crashes. Existing research by Novat et al. as well as Sinha et al. used these data to explore and identify crash patterns [15,16]. Lee et al., on the other hand, used 260 automated vehicles crash reports from California to examine automated vehicle crash patterns [17].

There have been already some studies that examined the injury severity of crashes that involved vehicles with automation. For the crash injury severity modeling, most of the existing studies focused on the ADS equipped vehicles and applied various modeling techniques including Bayesian inference [18], classification and regression tree model [19], XGBoost [20], decision trees and association rules [21], bootstrap-based logistic regression [22], and various other machine learning methods [23]. There are only a few studies that examined crash injury severity using the automated vehicle crash data. Some of the employed databases in past work included the NHTSA (associated with ADAS features) [24], Strategic Highway Research Program 2 (SHRP 2) Naturalistic Driving Study (NDS) (focused on drivers/operators) [25], and German In-depth Accident Study (GIDAS) dataset (related to pedestrians) [26]. Another popular approach of evaluating the role of driving automation on crash injury severity outcomes is simulation [27-29]. However, simulation techniques might not always reveal the real causality.

In addition to the crash severity analyses, other related topics were also examined, including crash type modeling [15,22,30,31], crash report text mining [32], Vulnerable Road Users' (VRU) involvement [33], risk/safety measurement methods [34], and disengagement analysis [35]. Although there exists a study that attempted to investigate the differences between crash injury outcomes for both ADS and ADAS equipped vehicles in rear-end crashes [36] as well as comparing the patterns and causes [37], the effects of the two different automation levels on injury severity still remain largely unexplored.

A summary of past research highlighting automated vehicle types, analytical methods, data sources, sample sizes, and whether the injury severity was considered are shown in Table 1.

Existing work that employed effective modeling and data pattern recognition was able to reveal several crash risk factors and provided insights for automated vehicle operations in practice. Past studies found that rear-end crashes was an important crash type for vehicles with ADS as well as different operation modes [38]. Ashraf et al. showed in their research that the majority of automated vehicle crashes were rear-end crashes and took place when the autonomous mode was engaged at road intersections [39]. Petrović et al. concluded that while the AV crashes involving conventional vehicles resulted in more rear-end crashes, they also yielded fewer pedestrian or broadside crashes, as well as aggressive driving behaviors [31]. Xu et al. performed more systematic crash analysis using the interaction data from connected and autonomous vehicles and identified the main crash patterns, which included the location type and driving mode [22].

Although some of the former studies have begun to examine the patterns and factors contributing to crash injury severity that relates to the vehicles equipped with driving automation [40], the research gap in this domain is still substantial. The knowledge regarding automated vehicle crashes remains fragmented, which stems primarily from the lack of data. Because of the data-related constraints, there has been no comprehensive study on the injury severity from crashes that involve vehicles with different driving automation capabilities. For example, limited research has been carried out to compare the crash patterns between the ADS and ADAS [41-43]. Most importantly, thus far, the existing data sources have not been completed, which in consequence has created a risk for unreliable analysis and incomplete policy recommendations. The current analysis aims to improve our understanding of the factors impacting crash injury severity relating to vehicles with automated driving capabilities of different levels.

Table 1. Summary of research related to the crashes of vehicles with driving automation.

| Reference number | Automation Type | Injury considered Yes (Y)/ No (N) | Method | Sample size |
|---|---|---|---|---|
| 21 | ADS | N (Probability of crash type) | BN model | 127 [2] |
| 22 | ADS | Y (Injured/ no injury) | Bagging | 259 [2] |
| 23 | ADS | Y (AV damage level) | Bayesian methods | 260 [2] |
| 25 | ADS | Y (KABCO) | CART and ordinal logistic regression | 107 [2] |
| 26 | ADS | Y (Injured/ no injury) | XGBoost and CART | 245 [2] |
| 27 | ADS | Y (1: more severe/ less severe crash) | Cost-sensitive classification and CART | 131 [2] |
| 28 | ADS | Y (Injured/ no injury) | Logistic regression model | 72 [2] |
| 29 | ADS | Y (Injured/ no injury) | SVM, CART, XGBoost, SHAP | 448 [5] |
| 30 | ADAS | Y (Potential crash prevention estimates) | Logistic regression model | 53 [2] |
| 31 | ADAS | N (Driver performance) | Multinomial logistic regression | 333 [2] |
| 32 | ADAS | Y (Pedestrian) | Injury severity score | 252 [2] |
| 33 | ADAS | Y (Complaints of pain) | Ordered probit model | 94 [2] |
| 34 | ADAS | Y (Injury reduction benefits) | Crash reconstruction and simulation | 198 [2] |
| 36 | ADS | N (Crash type) | SVM., NB, RF and NN | 19 (Google) |
| 37 | ADS | N (Crash type, driving errors) | Statistical analysis | 369 (SHRP2) |
| 38 | ADS | N (Text mining, crash pattern) | Bayesian network joint modeling method | 444 (GIDAS) |
| 39 | ADS | N (VRU involvement) | SVM, NB, RF, NN | 248 (Taxis) |
| 40 | ADS, ADAS (Simulation) | N (Risk measurement) | Statistics of sparse critical events | Microsimulation data (GIDAS) |
| 41 | ADS, ADAS | Y (Chance of rear-end crashes) | Binomial logistic regression | 214 [5] |
| 42 | ADS, ADAS | Y (Crash patterns, contact area, highest injury severity) | Logistic regression models | 202 valid data [5] |
| 44 | ADS | N (Crash reasoning) | Decision tree and association rule methods | 198 [5] |
| 45 | ADS | N (Crash involvement rates) | Statistical analysis | 19 (Google) |
| 46 | ADS, ADAS | N (ADAS/ADS test evaluation) | Optimal evaluation courses | Testing data collected by in-vehicle sensors |

[2] indicates CA DMV data source, [5] indicates NHTSA data source

# 3. DATA PREPARATION

## 3.1. DATA COLLECTION

The current study uses the NHTSA [5] and the CA DMV [2] crash databases. A total of 1,001 ADAS crash records were collected from NHTSA and 548 ADS crash records were collected from the CA DMV. The ADAS dataset corresponds to the crash records of the vehicles that are equipped with SAE Level 2 capabilities, while the ADS dataset includes information on crashes related to the vehicles with the automation systems of SAE Level 4.

The data were also examined in the context of crashworthiness. In essence, crashworthiness focuses on protecting occupants and minimizing fatal and serious injuries by implementing innovative vehicle designs, safety countermeasures, and equipment. Given the vehicles' similarity and to aid in model interpretation, crashworthiness was not included as an explanatory variable in the model.

**NHTSA crash database**

In June 2021, the NHTSA implemented a Standing General Order that requires manufacturers and operators to report crashes for vehicles that are equipped with ADS (SAE Level 4) or ADAS (SAE Level 2). Hence, the NHTSA crash database includes the information on the ADAS and ADS vehicle crashes from June 2021 that were collected from NHTSA's Crash Investigation Sampling System, Crash Report Supplement Datasheets, and Special Crash Investigations. Since crash records may be incomplete and unverified, NHTSA filtered and verified each crash record manually to prevent invalid, missing, or duplicated data. The data are available in the link provided in the reference section [5]. Entities listed in the NHTSA General Order are required to submit a crash report in cases when a crash within 30 seconds of Level 2 ADAS usage involves a vulnerable road user or leads to fatality, towing of a vehicle, airbag deployment, or hospitalization. Similarly, if there was any property damage or injury resulting from a crash within 30 seconds of the ADS equipped vehicle usage that takes place, entities mentioned in the General Order must also report it. While the data were comprehensive, the NHTSA dataset comes with several limitations. First there was variability in data access and telemetry across different vehicles. The capacity to record and transmit crash data significantly differs based on the manufacturer and the driving automation system. Vehicles equipped with ADS generally have advanced sensors and cameras for thorough data collection. Conversely, Level 2 (ADAS equipped) vehicles might have lacked robust data recording capabilities, potentially causing duplicated or incomplete information in the reported data. Second, there were incomplete and unverified incident report data. Entities responsible for reporting must submit incident reports within a specified timeframe. Consequently, these initial reports might have lacked comprehensive or verified information, such as surface conditions, passenger details, or injuries. Third, there was an issue of multiple reports for a single crash. A single crash that meets the reporting criteria could generate multiple reports, including an initial report, a follow-up at the 10-day mark, and subsequent updates. Different entities such as vehicle manufacturers, system developers/suppliers, and vehicle operators might all be obligated to report the same crash.

To validate the geospatial data, the focus was first on the incomplete data records. This involved checking the NHTSA reports to identify any missing incident records. Before any data imputation occurred, a manual review is conducted to rectify gaps and verify the accuracy of existing information. Next, duplicate entries utilizing geospatial and temporal analysis were addressed. Instances with the same location and time data, primarily due to shared fields like VIN, were identified and one of the duplicate records were removed. After the spatial-temporal filtering, the data were subjected to further verification using attributes such as vehicle ID, report ID, and same vehicle ID. In addition to the coordinates provided by the official crash reports, geospatial data were sourced from OpenStreetMap and cross-validated using Google Maps. The determination of location relied on both, directly available geospatial description and cross-referenced ones. The locations cross-referenced from the visual data were manually validated with significant landmarks visible in Google Maps' satellite images and street views of the crash scenes.

**CA DMV crash reports**

The other data source that was used in this study was the CA DMV OL316 Report of Traffic Collision Involving an Autonomous Vehicle. The report form also includes the details that are related to the crash type, damage, crash location, and injury severity, as well as the crash scene information such as the weather, lighting condition, surface condition, pre-crash conditions, together with other associated factors often used for the AV crash analysis. For the CA DMV, manufacturers conducting autonomous vehicle tests are obligated to report crashes resulting in property damage, bodily injury, or death within 10 days of the incident. The crash reports are available upon request, and the current analysis includes all the reported crashes from 2014 to March 2023 (the data were requested in March 2023 from the California Department of Motor Vehicles) [2].

Because the crash reports were provided in a pdf format and their design often varied by year, it would be cumbersome to manually retrieve the key information from the original reports. To overcome this issue, we employed the pdf parser and a Large Language Model [44] – ChatGPT 3.5 - to automatically extract crash attributes from the raw reports [45]. After the automated process was completed, manual checks and corrections were implemented for each report to ensure data quality and validity.

The abovementioned data sources were combined by matching the same or related attributes across the sources and organized to align with the NHTSA crash database format. Samples of data relating to the crash injury severities from the ADS and ADAS equipped vehicles were modeled separately. Furthermore, to contribute to the open-source research community, the processed and combined data were made available to the public and denoted as Autonomous Vehicle Operation Incident Dataset (AVOID). Detailed information about each step of data processing and gathering and their format is available in Zheng et al. [46].

## 3.2. DATA DESCRIPTION

To model the injury severity for vehicles with different automation levels, the dataset was divided into two categories. The first one included information on the crashes of the vehicles equipped with the ADAS (that still require a driver), while the second one contained information on crashes of vehicles equipped with the ADS (capable of automated driving). To ensure a comprehensive analysis, a multi-source dataset is used, which included 1,001 crashes of Level 2 vehicles (with ADAS) from the National Highway Traffic Safety Administration (NHTSA), and 548 crashes of Level 4 vehicles (with ADS) from the California Department of Motor Vehicles (CA DMV).

The final datasets contained various independent variables that are categorized into four major classes: road and environments characteristics (including the weather, road type, road description, surface condition, and lighting condition), pre-crash characteristics (including vehicle manufacture year, mileage, driver type, speed, and pre-crash movement), subject vehicle contact area, and crash details characteristics (including crash injury severity, airbag engagement, and vehicle towing information).

We have referred to the National Highway Traffic Safety Administration's classification for crash severity. This system delineates crashes into five categories: fatality, serious injury, moderate injury, minor injury, and no injuries. The definition of crash severity for SGO (Standing General Order on Crash Reporting) is consistent with the crash reports from CA DMV, which can be found in the KABCO injury classification scale and definitions: fatal injury, severe injury, other visible injury, and complaint of pain. From the crash reports in the CA DMV, vehicle damage and narratives were used to add the driver injury severity of crashes to both ADAS and ADS equipped vehicles. Lastly, three severity levels were used in the analysis: moderate/severe injury (including fatalities), minor injury, and no-injury. The first category, moderate/severe injury crashes, encompasses crashes where the involved individual sustains moderate to severe injuries and undergoes medical treatment, including those that lead to fatalities. The next category, minor injury crashes, refers to a crash where the involved individual might have incurred injuries but did not pursue medical care. The third category, no injury crashes, is applied to label the crashes where there are no reported or claimed injuries. A summary of the full sample of the available characteristics is presented in Table 2. Two kinds of drivers were considered: consumer and test. Consumers encompass individuals operating commercially available systems independently of manufacturers and are known as users or operators; and test drivers, who encompass individuals within the vehicle in a testing capacity or those providing remote assistance in commercial or testing scenarios.

A few interesting findings can be immediately noted by examining Table 2. For instance, while for the ADAS equipped vehicle crashes, over half of them took place on a highway (67 percent), most crashes involving the ADS equipped vehicles took place in more urban settings such as intersections (48 percent) and on local streets or roads (46 percent). Compared to vehicles with ADS, most highway crashes in the dataset involved the ADAS equipped vehicles. To some degree, it is expected that most crashes involving ADAS equipped vehicles take place on highways and most crashes related to the ADS equipped vehicles occur in other types of environments. These patterns likely reflect the limitations and design intentions of the respective systems. The ADAS is tailored for simpler highway driving conditions, while the ADS is generally intended to handle more complex environments, such as those encountered in urban settings. Therefore, the crash locations align with their respective systems' design intentions. In low-light conditions, such as darkness or dusk and dawn, the majority of crashes in the dataset are related to the ADAS equipped vehicles, as presented in Table 2.

Furthermore, the crash speed data shows obvious differences between the crashes involving the ADAS and ADS equipped vehicles. While 80 percent of crashes of vehicles equipped with ADS in the dataset take place at speeds equal or less to 20 mph, only 13 percent of crashes involving vehicles equipped with ADAS happened at the same speeds.

**Table 2.** Summary statistics for the variables included in the final model estimation.

| Variable type | Variable description | ADAS | | ADS | |
|---|---|---|---|---|---|
| | | Mean | Standard deviation | Mean | Standard deviation |
| **Road and environment** | | | | | |
| Weather | Clear sky indicator (1 if weather is clear, 0 otherwise) | 0.61 | 0.239 | 0.87 | 0.129 |
| | Cloudy sky indicator (1 if weather is cloudy, 0 otherwise) | 0.20 | 0.169 | 0.09 | 0.086 |
| | Rain indicator (1 if rain, 0 otherwise) | 0.18 | 0.142 | 0.03 | 0.045 |
| | Fog indicator (1 if there is fog, 0 otherwise) | 0.01 | 0.006 | 0.01 | 0.004 |
| Road type | Highway indicator (1 if a highway, 0 otherwise) | 0.67 | 0.223 | 0.02 | 0.020 |
| | Intersection indicator (1 if an intersection, 0 otherwise) | 0.13 | 0.112 | 0.48 | 0.250 |
| | Street indicator (1 if road is a local street, 0 otherwise) | 0.19 | 0.158 | 0.46 | 0.249 |
| | Parking lot indicator (1 if a parking lot, 0 otherwise) | 0.01 | 0.008 | 0.04 | 0.037 |
| Road description | No unusual conditions indicator (1 if there are no unusual conditions, 0 otherwise) | 0.90 | 0.091 | 0.97 | 0.027 |
| | Traffic incident/work zone indicator (1 if traffic incident/work zone, 0 otherwise) | 0.10 | 0.091 | 0.03 | 0.027 |
| Surface | Dry surface indicator (1 if surface is dry, 0 otherwise) | 0.82 | 0.146 | 0.89 | 0.090 |
| | Wet surface indicator (1 if surface is wet, 0 otherwise) | 0.17 | 0.140 | 0.09 | 0.072 |
| | Snow/slush/ice indicator (1 if surface is covered with snow/slush/ice, 0 otherwise) | 0.01 | 0.010 | 0.03 | 0.021 |
| Lighting | Dark– unlit indicator (1 if it is dark– not lit), 0 otherwise) | 0.09 | 0.081 | 0.01 | 0.004 |
| | Dark– lit indicator (1 if it is dark– lit, 0 otherwise) | 0.15 | 0.130 | 0.21 | 0.168 |
| | Daylight indicator (1 if daylight, 0 otherwise) | 0.72 | 0.205 | 0.75 | 0.186 |
| | Dawn/dusk indicator (1 if it is dawn/dusk, 0 otherwise) | 0.04 | 0.043 | 0.03 | 0.028 |
| **Pre-crash conditions** | | | | | |
| Manufacture year | Older vehicle indicator (1 if a vehicle was made before 2020, 0 otherwise) | 0.46 | 0.249 | 0.45 | 0.248 |
| | Newer vehicle indicator (1 if a vehicle was made in and after 2020, 0 otherwise) | 0.54 | 0.249 | 0.55 | 0.248 |
| Mileage | Mileage below or equal to 50,000 indicator (1 if a vehicle mileage is less than or equal to 50,000 miles, 0 otherwise) | 0.64 | 0.216 | 0.84 | 0.088 |
| | Mileage above 50,000 indicator (1 if a vehicle mileage is more than 50,000 miles, 0 otherwise) | 0.18 | 0.114 | 0.10 | 0.133 |
| | Unknown | 0.18 | 0.11 | 0.06 | 0.057 |
| Driver type | Consumer indicator (1 if driver type is consumer, 0 otherwise) | 0.98 | 0.013 | 0.18 | 0.147 |
| | Test indicator (1 if driver type is test, 0 | 0.02 | 0.017 | 0.82 | 0.147 |
| | otherwise) | | | | |
| Speed | Below or equal to 20 mph indicator (1 if speed is below or equal to 20 mph, 0 otherwise) | 0.13 | 0.111 | 0.80 | 0.147 |
| | More than 20mph and less than 40 mph indicator (1 if speed is more than 20 mph and less than 40 mph, 0 otherwise) | 0.36 | 0.141 | 0.08 | 0.055 |
| | More than 40 mph and less than 60 mph indicator (1 if speed is more than 40 mph and less than 60 mph, 0 otherwise) | 0.17 | 0.185 | 0.04 | 0.039 |
| | Above or equal to 60 mph indicator (1 if vehicle speed is above or equal to 60 mph, 0 otherwise) | 0.16 | 0.199 | 0.02 | 0.020 |
| | Unknown | 0.18 | 0.11 | 0.06 | 0.057 |
| Pre-crash movement | Proceeding straight indicator (1 if pre-crash movement is proceeding straight, 0 otherwise) | 0.71 | 0.211 | 0.46 | 0.248 |
| | Merging indicator (1 if pre-crash movement is merging, 0 otherwise) | 0.02 | 0.021 | 0.09 | 0.082 |
| | Crossing roadway indicator (1 if pre-crash movement is crossing roadway, 0 otherwise) | 0.02 | 0.021 | 0.01 | 0.004 |
| | Turn indicator (1 if pre-crash movement is turn, 0 otherwise) | 0.08 | 0.068 | 0.11 | 0.112 |
| | Changing lanes indicator (1 if pre-crash movement is changing lanes, 0 otherwise) | 0.04 | 0.045 | 0.15 | 0.082 |
| | Stopping indicator (1 if pre-crash movement is stopping, 0 otherwise) | 0.13 | 0.120 | 0.18 | 0.130 |
| Crash type | Fixed object indicator (1 if crash with fixed object, 0 otherwise) | 0.27 | 0.198 | 0.02 | 0.018 |
| | Vehicle indicator (1 if crash with another vehicle, 0 otherwise) | 0.58 | 0.246 | 0.75 | 0.188 |
| | Truck and van indicator (1 if crash with truck and van, 0 otherwise) | 0.12 | 0.109 | 0.12 | 0.047 |
| | Motorcycle indicator (1 if crash with motorcycle, 0 otherwise) | 0.01 | 0.013 | 0.06 | 0.109 |
| | Pedestrian indicator (1 if crash with pedestrian, 0 otherwise) | 0.02 | 0.024 | 0.05 | 0.057 |
| **AV Vehicle contact area indicator** | | | | | |
| | Left impact indicator (1 if subject vehicle contact area is left, 0 otherwise) | 0.08 | 0.130 | 0.12 | 0.105 |
| | Right impact indicator (1 if subject vehicle contact area is right, 0 otherwise) | 0.07 | 0.139 | 0.14 | 0.120 |
| | Rear impact indicator (1 if subject vehicle contact area is rear, 0 otherwise) | 0.52 | 0.222 | 0.45 | 0.249 |
| | Front impact indicator (1 if subject vehicle contact area is front, 0 otherwise) | 0.28 | 0.220 | 0.23 | 0.179 |
| | Top or bottom impact indicator (1 if subject vehicle contact area is top or bottom, 0 otherwise) | 0.05 | 0.022 | 0.06 | 0.047 |
| **Crash details** | | | | | |
| Severity | No-injury indicator (1 if injury severity outcome is no injury, 0 otherwise) | 0.89 | 0.104 | 0.66 | 0.241 |
| | Minor injury indicator (1 if injury severity outcome is minor, 0 otherwise) | 0.05 | 0.045 | 0.26 | 0.180 |
| | Moderate injury and higher indicator (1 if | 0.06 | 0.057 | 0.08 | 0.139 |
| | injury severity outcome is moderate and higher including fatality, 0 otherwise) | | | | |
| Airbags | Airbags engaged indicator (1 if airbags are engaged, 0 otherwise) | 0.08 | 0.072 | 0.05 | 0.045 |
| | Airbags not engaged indicator (1 if airbags are not engaged, 0 otherwise) | 0.92 | 0.072 | 0.95 | 0.045 |
| Towed | Towed vehicle indicator (1 if vehicle is towed, 0 otherwise) | 0.08 | 0.073 | 0.03 | 0.018 |
| | Vehicle not towed (1 if vehicle is not towed, 0 otherwise) | 0.92 | 0.073 | 0.97 | 0.018 |

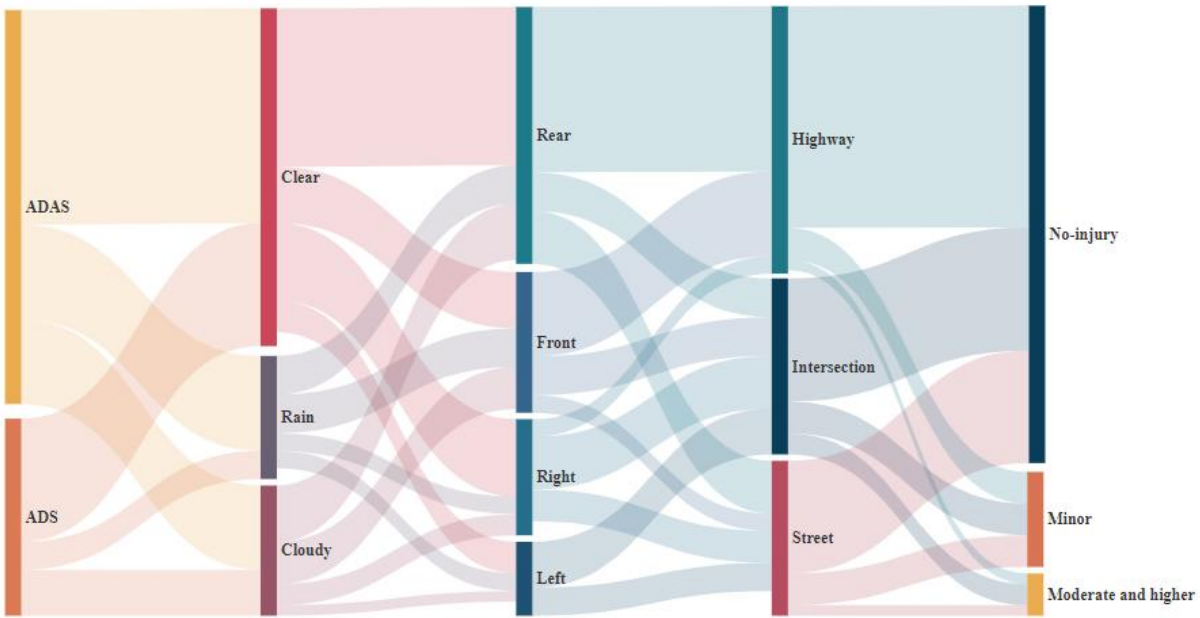

**Figure 1.** Relationship between the factors influencing injury severity.

Figure 1 presents the associations between the vehicle types, crash features, and injury severity levels. The width of each band corresponds to the proportion of the previous layer that contributes to the next layer. It can be concluded that most crashes in the entire dataset (ADAS and ADS) took place in clear weather conditions, at intersections or highways and impacted the rear of the vehicle.

## 4. METHODOLOGY

To gain more insights into the factors that play a role in the injury severity level, two random parameters multinomial logit models with heterogeneity in the means and variances were estimated. In each model, three injury severity categories were considered: no injury, minor injury, and moderate/severe injury (including fatalities). The chosen modeling approach allows the mean and variance of random parameters to be functions of explanatory variables and thus provides additional accuracy in studying unobserved heterogeneity. Random parameter ordered probit model was also estimated; however, it offered less interpretability and a worse fit. The multinomial logit model with heterogeneity in the means and variances has been widely applied in traffic safety analysis [47,48]. First, a function that determines injury severity probabilities is defined as,

$$F_{kn} = \boldsymbol{\beta}_k \boldsymbol{X}_{kn} + \varepsilon_{kn}, \qquad (1)$$

where $F_{kn}$ is the function determining the probability of injury-severity category $k$ in crash $n$, $\boldsymbol{X}_{kn}$ is the vector of explanatory variables that affect motorcyclist injury-severity level $k$, $\boldsymbol{\beta}_k$ is the vector of estimable parameters, and $\varepsilon_{kn}$ is the error term which is assumed to be a generalized extreme value distributed [49]. Given that, a random parameter multinomial logit model of injury severity probabilities can be derived as [50]:

$$P_n(k) = \int \frac{EXP(\boldsymbol{\beta}_k \boldsymbol{X}_{kn})}{\sum_{\forall K} EXP(\boldsymbol{\beta}_k \boldsymbol{X}_{kn})} f(\boldsymbol{\beta}_k|\boldsymbol{\varphi}_k) d\boldsymbol{\beta}_k, \qquad (2)$$

where $P_n(k)$ is the probability that crash $n$ results in injury category $k$, $f(\boldsymbol{\beta}_k|\boldsymbol{\varphi}_k)$ is the density function of $\boldsymbol{\beta}_k$ and $\boldsymbol{\varphi}_k$ refers to the vector of parameters describing the mixing density function (mean and variance), with all the other terms being previously defined. To give more flexibility in accounting for unobserved heterogeneity, with the mixing distribution now allowing parameters to vary across observations $n$, the $\boldsymbol{\beta}_{kn}$ vector can be made to be a function of variables that affect its mean and variance as [51,52]:

$$\boldsymbol{\beta}_{kn} = \beta_k + \Theta_{kn} \boldsymbol{Z}_{kn} + \sigma_{kn} EXP(\Psi_{kn} \boldsymbol{W}_{kn}) v_{kn} \qquad (3)$$

where $\boldsymbol{\beta}_k$ is the mean parameter estimate across all crashes, $\mathbf{Z}_{kn}$ is a vector of observation-specific explanatory variables that captures heterogeneity in the mean that affects injury-severity level alternative $k$, $\boldsymbol{\Theta}_{kn}$ is a corresponding vector of estimable parameters, $\mathbf{W}_{kn}$ is a vector of observation-specific explanatory variables that captures heterogeneity in the standard deviation (variance) $\sigma_{kn}$ with corresponding parameter vector $\Psi_{kn}$ and $v_{kn}$ is a disturbance term.

The final model estimation was performed by simulated maximum likelihood approach with 1,000 Halton draws as they were found to deliver more efficient distribution of simulation draws than purely random draws [47]. Although numerous density functions can be assumed for $f(\boldsymbol{\beta}_k|\boldsymbol{\varphi}_k)$ (Equation 2) normal distribution was assumed for random parameters as it was determined to be statistically superior, which is also consistent with past work [52,53].

Although extensive testing has been done, no heterogeneity in the variance was detected in the models and only one model had statistically significant heterogeneity in the mean.

To evaluate the model transferability (between the ADAS and ADS crash severity injury outcomes), a series of likelihood ratio tests were performed. The test statistic for this can be defined as follows [50,54]:

$$\chi^2 = -2[\text{LL}(\beta_{ba}) - \text{LL}(\beta_a)] \qquad (4)$$

where $LL(\beta_{ba})$ represents the converged log-likelihood for a model that incorporates parameters from the $b$ type vehicle type on crash data from the vehicle $a$ data subset. Similarly, $LL(\beta_a)$ stands for the converged log-likelihood of a model that uses crash data exclusively from the $a$ vehicle type subset (in this case ADS and ADAS equipped vehicles). The test can be reversed using $LL(\beta_{ab})$ and $LL(\beta_b)$. The statistic is $\chi^2$ and the number of degrees of freedom for this test is equal to the number of parameters being estimated [55].

Alternatively, the transferability between the combined model and the individual models was also evaluated using an additional likelihood ratio test. The test statistic can be described as [53,56]:

$$\chi^2 = -2[LL(\beta_T) - LL(\beta_a) - LL(\beta_b)] \qquad (5)$$

where the log-likelihood value at the point of model convergence when using all accessible crash data records from both ADAS and ADS equipped vehicles is represented as $LL(\beta_T)$. The $LL(\beta_a)$ and $LL(\beta_b)$ indicate the log-likelihood value at the convergence point for the models that solely use the crash data for each vehicle type. The $X^2$ statistic is $\chi^2$ distributed with degrees of freedom equal to the summation of the number of estimated parameters in both models (crash severity outcomes of the ADS and ADAS equipped vehicles) minus the number of estimated parameters in the overall model. The results of the $X^2$ statistic give the confidence level that the null hypothesis (that the parameters are the same) can be rejected. [55].

To support the interpretation of the model estimation findings, marginal effects were calculated. Marginal effects capture the effect that a one-unit change in an explanatory variable has on the probability of an injury-severity outcome [57]. The values of the corresponding marginal effects were calculated for each observation and then they were averaged over the population of observations. For variables with mixed effects, which are the variables with significant parameters in more than one injury severity function, net marginal effects are reported.

## 5. RESULTS

The estimation results for the crash injury severity involving the ADAS equipped vehicles (SAE Level 2) are presented and followed by the results of the ADS equipped vehicles (SAE Level 4). Because the ADAS and ADS are two different levels of driving automation, their crash reporting is based on different reporting criteria. Since they likely exhibit different behaviors in traffic, a series of the likelihood ratio tests were performed to ensure and validate the need for two separate models. The results of the likelihood ratio tests rejected the null hypothesis that both models are the same with over 99% confidence, justifying the necessity for estimating distinct models for each vehicle type. Tables 3 and 4 present the test results and therefore two separate estimations are presented for injury severity outcomes for vehicles with different levels of automation in Table 5 and Table 6. One random parameters multinomial logit model was estimated for the ADAS crash injury data (Table 5) and one random parameters multinomial logit model (Table 6) was estimated for the ADS crash injury data.

**Table 3.** The results of likelihood ratio tests between different vehicle types based on random parameters approaches with heterogeneity in the means and variances.

| $a$ | $b$ | |
|---|---|---|
| | ADAS | ADS |
| ADAS | – | 85.19 (17) [>99.99%] |
| ADS | 30.94 (13) [>99.99%] | – |

**Table 4.** Likelihood ratio tests results for the joint model (the number of parameters in each model (Equation 5)

| LL(overall) | LL(*ADAS*) | LL(ADS) | $\chi^2$ | Degrees of freedom | Confidence level |
|---|---|---|---|---|---|
| -322.94 (26) | -161.49 (13) | -220.49 (17) | 118.08 | 4 | >99.99% |

Likelihood ratio tests were performed to ensure that only the statistically superior results are presented. As confirmed by the likelihood ratio tests, the random parameter models were indeed statistically superior at 99% confidence level (the number of degrees of freedom was equal to the difference between the number of parameters being estimated in the fixed and random parameter models). A total of 13 variables were found significant in the model examining crash injury severity involving the ADAS (SAE Level 2) equipped vehicles (Table 5), and a total of 17 variables were found significant in the model analyzing crash injury severity involving ADS (SAE Level 4) equipped vehicles (Table 6).

## 5.1 RESULTS: SELECT FINDINGS ON CRASH INJURY SEVERITY OUTCOMES INVOLVING ADAS (SAE Level 2) EQUIPPED VEHICLES

The statistically significant variables found in the estimated model were divided into several categories: random parameters, road and environment factors, crash details, pre-crash conditions, and lastly variables with mixed effects (Table 5). The model performance is very good as indicated by the $\rho^2$ of 0.82.

Considering the random parameter multinomial logit model's results relating to the crash injury severity involving vehicles equipped with ADAS, the front impact indicator was found to be a random parameter in the model (Table 5) and decrease the probability of a minor injury. The model estimation results emphasize the complex and heterogenous relationship between the injury severity outcomes and ADAS equipped vehicles with front impact damage. While the current analysis is exploratory, future work might consider analyzing the front impact damage and relating it to a particular vehicle combination (ADAS lead or following).

Turning to the fixed parameters that were found statistically significant in the model (Table 5), dry surface indicator was found to decrease the probability of a moderate/severe injury for the ADAS equipped vehicles relative to other surface conditions. According to the model estimation results, the model revealed that the dry surface indicator decreases the probability of a moderate/severe injury by 0.0221 (as indicated by the marginal effects in Table 5). The condition of the surface is likely related to the friction between the tires and the road and the findings reflect its overall effect on the injury severity outcomes.

Considering crash details, according to the model results older vehicle indicator (made before 2020) increases the probability of a minor injury by 0.0111. Contact area, on the other hand, such as the rear of the vehicle indicator was found to decrease the likelihood of a moderate/severe injury while simultaneously increasing the probability of the alternatives (as indicated by the marginal effects in Table 5).

Turning to the pre-crash conditions, driver type indicators were found statistically significant in the model. The test driver type indicator in the ADAS model was found to increase the probability of a minor injury by 0.004 as suggested by the marginal effects in Table 5. The model results indicate that the operation mode and driver type indicators play a considerable role in crash injury severity outcomes. While rear-end crashes are generally considered to be of lower severity compared to other crash types [59], our dataset reveals a notable trend: a significant proportion of ADAS involved crashes on highways (68 percent) are rear end, where vehicles typically travel at higher speeds. It was also found that the rear impact indicator decreases the probability of a moderate/severe injury by 0.003.

When considering the variables with mixed effects, the mileage less than 50,000 miles indicator and the airbag engagement have shown to be particularly complex as these indicators were found to be significant parameters in more than one injury severity alternative, which led to reporting the net marginal effects. According to the model estimation results (Table 5), mileage less than 50,000 miles indicator was found to decrease the probability of a minor and moderate/severe injury. Airbag engagement indicator also resulted in a lower probability of both minor and moderate/severe injury for vehicles equipped with SAE Level 2. Airbag related findings results are somewhat intuitive, as they signal their effectiveness in decreasing the injury severities.

**Table 5.** Random parameters logit model on crash-injury severity for vehicles equipped with **ADAS (SAE Level 2)** (parameters defined for [N] No injury; [M] Minor injury; [MM] Moderate/severe injury).

| Variable | Estimated parameter | t-statistic | Marginal effects | | |
|---|---|---|---|---|---|
| | | | No injury | Minor injury | Moderate/severe injury |
| Constant [M] | -1.34 | -2.11 | | | |
| Constant [MM] | 5.61 | 3.34 | | | |
| *Random parameters* | | | | | |
| Front impact indicator (1 if subject vehicle contact area of vehicle is front, 0 otherwise) [M] | 1.64 | 1.88 | -0.0081 | -0.0006 | 0.0087 |
| *Standard deviation of parameter distribution* | *1.85* | *1.88* | | | |
| *Road and environment* | | | | | |
| Dry surface indicator (1 if surface was dry, 0 otherwise) [MM] | -6.15 | -5.99 | 0.0188 | 0.0033 | -0.0221 |
| *Crash details* | | | | | |
| Older vehicle indicator (1 if a vehicle was made before 2020, 0 otherwise) [M] | 0.66 | 1.72 | -0.0105 | 0.0111 | -0.0006 |
| Fixed object indicator (1 if vehicle crash with fixed object, 0 otherwise) [M] | 0.83 | 2.20 | -0.0166 | 0.0174 | -0.0008 |
| Rear indicator (1 if subjective vehicle contact area is rear, 0 otherwise) [MM] | -2.36 | -1.67 | 0.0026 | 0.0004 | -0.0030 |
| *Pre-crash conditions* | | | | | |
| Test indicator (1 if driver type is test, 0 otherwise) [M] | 2.08 | 2.50 | -0.0036 | 0.0040 | -0.0004 |
| ***Variables with mixed effects (variables with significant parameters in more than one injury severity function, net marginal effects reported)*** | | | | | |
| Mileage is less than 50,000 miles indicator (1 if a vehicle mileage is less than 50,000 miles, 0 otherwise) [M] | -1.54 | -3.57 | 0.0593 | -0.0424 | -0.0169 |
| Mileage is less than 50,000 miles indicator (1 if a vehicle mileage is less than 50,000 miles, 0 otherwise) [MM] | -7.64 | -5.24 | | | |
| Airbag engagement indicator (1 if airbags were engaged, 0 otherwise) [M] | -1.29 | -2.39 | 0.0669 | -0.0425 | -0.0244 |
| Airbag engagement indicator (1 if airbags were engaged, 0 otherwise) [MM] | -2.93 | -2.13 | | | |
| Number of observations | | | 1,001* | | |
| Log likelihood at zero, LL (0) | | | -896.47 | | |
| Log likelihood at convergence, LL(β) | | | -161.49 | | |
| $\rho^2 = 1 - LL(\beta)/LL(0)$ | | | 0.82 | | |

*185 observations were removed due to missing values

## 5.2 RESULTS: SELECT FINDINGS ON CRASH INJURY SEVERITY OUTCOMES INVOLVING ADS (SAE Level 4) EQUIPPED VEHICLES

Table 6 presents the results of the random parameters logit model with heterogeneity in the mean of random parameter on crash-injury severity for vehicles equipped with ADS. The variables found statistically significant in the model were divided into the following categories: random parameters, road and environment factors, crash details, pre-crash conditions, and lastly variables with mixed effects (Table 6). The model performance is excellent as indicated by the $\rho^2$ of 0.61.

Turning to the random parameter category, the intersection indicator produced statistically significant random parameter with statistically significant heterogeneity in the mean captured by the cloudy sky indicator (Table 6). Interestingly, the intersection indicator was found to increase the probability of a moderate or severe injury (while simultaneously decreasing the probability of the other alternatives). The results reveal not only the varied impact of the intersection environment on the injury severity outcomes but also suggest the importance of the visibility and weather/lighting conditions (such as glare or lack of thereof) and the role they play in the crash outcomes.

The road and environmental factors such as the weather conditions at the time of a crash were found to also be important in the injury severity outcomes (namely rainy weather indicator and cloudy sky indicator). While former work, to some degree, stressed out the necessity to consider ambient conditions and how they effect sensor performance, the current findings relate these factors to crash injury outcomes [58, 60] (Table 6).

Turning to the crash details, the location of the impact was determined to be statistically significant in the model with front impact collisions increasing the probability of a minor injury as a result of the crash by 0.0478. The rear impact indicator decreased the probability of minor injuries in crashes involving the ADS equipped vehicles. While rear-end crashes are generally considered to be of lower severity compared to the other crash types [59], the current dataset reveals that only 0.15 percent of rear-end crashes involving ADS took place on highways. This low percentage suggests that highways constitute a smaller segment of the overall environments where ADS vehicles have been commonly operated. Considering the driver type indicator, it was also statistically significant and increased the probability of a minor injury. In the model, the marginal effects for the test drive type indicator indicated an increase in the probability of a minor injury by 0.025 relative to other driver types (see marginal effects in Table 6).

Turning to the other pre-crash conditions such as movement and speed, the findings related to the ADS equipped vehicles reveal that while turning and changing lanes indicators increase the probability of a minor injury by 0.0223 and 0.0210, respectively, stopping indicator decreases the probability of a moderate/severe injury. It is likely because stopping often involves more deliberate and regulated vehicle movements, and of course decreasing speed [61]. Speed-related findings for the ADS equipped vehicles are also worth noting. While the median speed variable indicating that the speed was above or equal to 20 mph and less than 40 mph decreased the probability of a minor injury, the median speed of above or equal to 40 mph and less than 60 mph indicator increased the probability of a moderate/severe injury. Exploring the injury severity outcomes at different speeds is essential to deepening the knowledge regarding the safety of vehicles with automation. Because in most cases ADS equipped vehicles use sensors and cameras that aid in detecting a range of objects, human related factors might have a decreased impact in the future. Nevertheless, these systems provide valuable assistance with steering, acceleration, and braking [62] and while they might not perfect, they do offer a promising value to improve safety outcomes.

Traffic incident/work zone indicator was found to produce statistically significant parameters in more than one injury severity alternative. This indicator was found to increase the probability of minor and moderate/severe injuries (as indicated by the net marginal effects in Table 6). Remarkably, the model estimation results suggest that reevaluating the effectiveness of the ADS systems in more complex environments where the geometry and other components are not standard is essential in improving the overall safety of these vehicles. While the number of crashes in incident/work zones in the current sample is limited, it is recommended that this type of crashes should be explored further once more data become available.

**Table 6.** Random parameter logit model with heterogeneity in the mean of random parameter on crash-injury severity for vehicles equipped with **ADS (SAE Level 4)** (parameters defined for [N] No injury; [M] Minor injury; [MM] Moderate/severe injury).

| Variable | Estimated parameter | t-statistic | Marginal effects | | |
|---|---|---|---|---|---|
| | | | No injury | Minor injury | Moderate/ severe injury |
| Constant [M] | -1.29 | -5.33 | | | |
| Constant [MM] | 2.01 | 4.11 | | | |
| *Random parameters* | | | | | |
| Intersection indicator (1 if an intersection, 0 otherwise) [MM] | -2.44 | -1.94 | -0.0276 | -0.0048 | 0.0324 |
| *Standard deviation of parameter distribution* | *4.32* | *3.74* | | | |
| *Heterogeneity in the mean of the random parameter* | | | | | |
| Intersection indicator: Cloudy indicator (1 if cloudy, 0 otherwise) | 6.20 | 2.57 | | | |
| *Road and environment* | | | | | |
| Rain indicator (1 if crash were rain, 0 otherwise) [M] | 1.43 | 1.79 | -0.0031 | 0.0054 | -0.0023 |
| Cloudy weather indicator (1 if the weather is cloudy, 0 otherwise) [MM] | 1.65 | 1.93 | -0.0062 | -0.0039 | 0.0101 |
| *Crash details* | | | | | |
| Rear impact indicator (1 if subject vehicle contact area is rear, 0 otherwise) [M] | -2.56 | -5.64 | 0.0295 | -0.0331 | 0.0036 |
| Front impact indicator (1 if subject vehicle contact area is front, 0 otherwise) [M] | 4.73 | 6.60 | -0.029 | 0.0478 | -0.0188 |
| Test type indicator (1 if driver type is test, 0 otherwise) [M] | -6.28 | -7.16 | 0.093 | 0.025 | -0.118 |
| *Pre-crash conditions* | | | | | |
| Turn indicator (1 if pre-crash movement is turn, 0 otherwise) [M] | 1.72 | 3.94 | -0.0177 | 0.0223 | -0.0046 |
| Changing lanes indicator (1 if pre-crash movement is changing lanes, 0 otherwise) [M] | 1.22 | 1.93 | -0.0081 | 0.021 | -0.0129 |
| Stopped indicator (1 if vehicle is stopped, 0 | -3.23 | -3.60 | 0.0078 | 0.0036 | -0.0114 |
| otherwise) [MM] | | | | | |
| Above or equal to 20 mph and less than 40 mph indicator (1 if speed is above or equal to 20 mph and less than 40 mph, 0 otherwise) [M] | -5.44 | -5.55 | 0.0193 | -0.0297 | 0.0104 |
| Above or equal to 40 mph and less than 60 mph indicator (1 if speed is above or equal to 40 mph and less than 60 mph, 0 otherwise)[MM] | 4.89 | 3.76 | -0.0075 | -0.0026 | 0.0101 |
| ***Variables with mixed effects (variables with significant parameters in more than one injury severity function, net marginal effects reported)*** | | | | | |
| Traffic incident/work zone indicator (1 if traffic incident in work zone, 0 otherwise) [M] | 3.5 | 2.72 | -0.0315 | 0.0106 | 0.0209 |
| Traffic incident/work zone indicator (1 if traffic incident/work zone, 0 otherwise) [MM] | 6.55 | 1.41 | | | |
| Number of observations | | | 548* | | |
| Log likelihood at zero, LL(0) | | | -565.79 | | |
| Log likelihood at convergence, LL(β) | | | -220.49 | | |
| $\rho^2 = 1-LL(\beta)/LL(0)$ | | | 0.61 | | |

*34 observations were removed due to missing values

## 6. SUMMARY AND CONCLUSIONS

This study conducted an exploratory analysis using a multi-source dataset (for more details regarding the dataset please refer to Zheng et al.[46]), which consists of crash reports for vehicles with different levels of automation (SAE Level 2 and Level 4). To ensure the reliability of the analysis, 1,001 (ADAS) and 548 (ADS) crash records were collected, checked and processed. One random parameters logit model and one random parameters logit model with heterogeneity in the means were estimated to understand the differences between the crash injury outcomes for ADAS and ADS crash data, respectively. Four types of variables were considered, and they included the subject vehicle contact area, road and environment, pre-crash movement, and crash details. Given the available variables, three levels of injury severity for the model estimation were used (no-injury, minor injury, and moderate/severe injury, including fatalities).

From the descriptive analysis it was found that while over half of the crashes (67 percent), involving the ADAS equipped vehicles (SAE Level 2, driver required) took place on a highway, a large number of crashes involving ADS equipped vehicles (SAE Level 4, no driver required) took place in more urban settings such as intersections (48 percent) and on local street or roads (46 percent).

Turning to the model estimation results, the dry surface indicator was particularly important and resulted in a 0.0221 lower probability of a moderate/severe injury for vehicles equipped with the ADAS. Concerning the road type, crashes occurring with test drivers were found to result in a higher probability of minor injuries for the vehicles equipped with the ADAS (Table 5). Traffic incident/work zone indicator resulted in an increased probability of a minor and moderate/severe injuries that involve the ADS equipped vehicles (SAE Level 4) (as suggested by the net marginal effects in Table 6). Because traffic work zones and incident zones can lead to sudden and abrupt modifications to the driving environment [63], special attention has to be paid to the performance of the ADS equipped vehicles in work zones in the future once more data become available. While the functioning of the ADS equipped vehicles in work zones has been identified to be particularly important in crash injury severity, a crash with fixed object indicator resulted in a 0.0008 lower probability of a moderate/severe injury for the ADAS equipped vehicles (as indicated by the net marginal effects in Table 5).

Concerning the road type, crashes occurring at intersections were found to have a higher probability of minor injuries for the vehicles equipped with the ADAS (Table 5). This finding indicates that the injury severity outcome will likely vary by the level of automation and environment.

Nevertheless, it is essential to recognize that Level 2 automation is designed to supplement the driver's input, not to take over the full spectrum of driving tasks. These systems significantly aid the driver yet require the driver to remain engaged and vigilant, particularly at intersections where visibility issues can arise. This understanding underlines the importance of the driver's active participation in conjunction with Level 2 automation to ensure safe operations. The responsibility for hazard perception at such levels of automation remains with the driver. Furthermore, turning or merging vehicles are frequently present at intersections [64], which can be challenging for the ADAS equipped cars to detect and respond swiftly. Although more research is needed to test the existing systems, the current analysis points to the variation in the injury severity between the ADAS and ADS equipped vehicles. The vehicles equipped with the ADAS still depend on human drivers to maneuver intersections, while the ADS equipped vehicles are engineered to function autonomously, relying on sensors and algorithms to detect and react to the surroundings [65]. Furthermore, the fact that the operator is aware of these systems can change their behavior and their level of alertness.

Studying the outcomes of injury severity for different drive type indicators and different levels of automation allows to further improve and optimize the operations and safety. Because the testing vehicles used for evaluating the ADAS are usually equipped with at least two primary control functions that work in unison (e.g., adaptive cruise control and lane centering), it is particularly important to evaluate new technological capabilities in the context of injury severity outcomes. Furthermore, testing vehicles often have an experienced operator who is very familiar with the technology and can react in real time to any unsafe conditions [44,66].

Interestingly, in the ADAS model, mileage is less than 50,000 miles was found to decrease the probability of both minor and moderate/severe injury. These findings could indicate that relatively small updates to the technology can result in significant shifts in the injury severity outcomes (under the assumption that vehicles with lower milage tend to be newer). Lastly, the location of impact has been especially important in this analysis, as it was found that the rear and front impact indicators increase the probability of a moderate/severe injury in the ADAS model. These findings suggest that designing vehicles with automated capabilities should continue to account for human drivers in the fleet as the sudden and unexpected braking done particularly by the ADAS equipped vehicles can cause a following vehicle to collide with its rear. To reduce the injury severity, manufactures may consider Automatic Emergency Braking (AEB), Blind Spot Monitoring (BSM), Lane Departure Warning (LDW), or Rear Cross-Traffic Alert (RCTA) [67]. Secondly, sensors can be integrated to detect road condition and enhance the performance of ADS and ADAS technologies [68].

Substantial differences were found between the ADS and ADAS systems and therefore the policy recommendations should target each system separately. For the ADAS equipped vehicles, further testing should evaluate situations with different areas of impact as well as during diverse road surface conditions. For the ADS equipped cars, more investments should be made in testing their operations at higher speeds, during more complex maneuvers such as changing lanes or turning, and in rainy weather.

Although significant insights into the factors affecting the crash injury severity of vehicles capable of automated driving have been obtained by analyzing real-world and multi-source data, the current research has its limitations. One of the main drawbacks of this study is the limited number of observations indicating higher levels of crash injury severity. An alternative approach would be to use a binary approach (injury and no injury) or nested logit model, however, some important pieces of information would be lost. It was still valuable to include all crash injury severities. Another limitation of this study is the exclusion of certain critical variables such as traffic control devices (e.g., stop signs, priority signs, and traffic lights) due to data unavailability. Lastly, the crash sample uses crashes over a period of time and therefore in the future it is important to capture temporal shifts relating to technology advancements and crash injury outcomes. As the data becomes available, it becomes increasingly critical to examine the true benefits of automated vehicles and compare their results with human-driven vehicles in real-world conditions.

## 7. AUTHOR CONTRIBUTIONS

S.D., M.A., D.W., O.Z., and Z.W. conceived the study. S.D., M.A., and N.B. wrote the manuscript. S.D. and N.B. estimated the models and conducted the analysis. M.A. and N.B. supervised the analysis and edited the manuscript.

## 8. COMPETING INTERESTS

The authors declare no competing interests.

## 9. DATA AND CODE AVAILABILITY

The data are open source and available at [https://github.com/UCF-SST-Lab/AVOID-Autonomous-Vehicle-Operation-Incident-Dataset.](https://github.com/UCF-SST-Lab/AVOID-Autonomous-Vehicle-Operation-Incident-Dataset) The codes for data validation and processing are available in the published GitHub repository: [https://github.com/UCF-SST-Lab/AVOID-Autonomous-Vehicle-Operation-Incident-Dataset.](https://github.com/UCF-SST-Lab/AVOID-Autonomous-Vehicle-Operation-Incident-Dataset) The quick tutorial and README file are also included in the repository for reference. Python scripts for geospatial data processing are prepared with the OSMnX package and offered in the repository, which can be referred to in the file Address2OSM.ipynb under the folder of code. Any updates will be published on GitHub.